\documentclass[cits]{PoS}

\title{Chemical tagging and the second r-process.}

\ShortTitle{The second r-process}

\author{\speaker{Camilla Juul Hansen}\thanks{Thanks to M. Bergemann, F. Primas, H. Hartman, K.-L. Kratz, S. Wanajo, N. Christlieb, O. Hallmann, B. Leibundgut, H. Nilsson, and NIC organizers.}\\
        Heidelberg University, ZAH, LSW\\
        E-mail: \email{cjhansen@lsw.uni-heidelberg.de}}

\abstract{Elements in the range 37 < Z < 47 provide key information on their
formation process. Several studies have shown that some of these elements
are formed by an r-process, that differs from the main r-process creating
europium. Through a detailed abundance study of Rb - Ag I will show, by
comparing these abundances to those of Ba and Eu, that their formation
processes differ. The formation process of Pd and Ag deviates from the
weak/main s-process as well as from the main r-process. Hence, Pd and Ag -
and to some extend Zr - are created by a second/weak r-process. However,
the characteristics and formation site of this process is not well
understood. The abundance ratios of Rb/Zr help constrain the neutron
number density of the formation site, while comparing the Pd and Ag
abundances to yield predictions can provide limitations on the entropy and
electron fraction of the formation environment. This study presents clues
on the second r-process. Furthermore, the formation processes of the heavy
elements might not differ in a clear cut way. Several of these
neutron-capture processes might yield various amounts of heavy elements
(e.g. Sr and Ba) at the same time or metallicity. This could possibly help
explain the large star-to-star abundance scatter for these two elements
below [Fe/H]= -2.5. Knowing their origin is important in the era of large
surveys (e.g Gaia-ESO). Strontium and barium will, limited by resolution
and signal-to-noise ratio, be the only detectable heavy elements in the
most metal-poor stars. Hence, they will, depending on metallicity, be the
main tracers of the weak and main s-/r-processes. Understanding the effects
of stellar parameters, synthetic spectrum codes, model atmospheres, and
NLTE on the Sr abundances are crucial to describe the chemical evolution
of our Galaxy. I will present these effects for Sr.}

\FullConference{XII International Symposium on Nuclei in the Cosmos,\\
		August 5-12, 2012\\
		Cairns, Australia}

\begin{document}

\section{Introduction}
The formation processes of neutron-capture elements are in many cases not well known. To date we have observational and theoretical evidence for the existence of two s-processes (weak and main), and one main r-process. Here observational indications for the presence of a second r-process will be shown (for details see \pos{[8]}). 
The indications will be made clear through a comparison of observationally derived stellar abundances of seven elements belonging to the four different formation processes.

The precision with which we can determine stellar abundances is higher than what we gain from model yield predictions and therefore also from Galactic chemical evolution (GCE) models. This means that we may use stellar abundances to constrain and improve yield predictions and GCE models. When applying the observationally derived abundances in this way, it is important that we know them very accurately, and that we understand how e.g. NLTE effects will affect the abundances of the heavy elements at all metallicities. 

Here I will focus on how the formation process of five neutron-capture elements (Sr, Y , Pd, Ag, and Ba) differ or agree, and how one of these elements (Sr) behaves under LTE and NLTE as a function of [Fe/H] (for details see \pos{[8]} and Hansen et al, 2012b accepted).

\section{Stellar abundances}
\begin{figure}[!h]
    \begin{center}
    \includegraphics[width=0.67\textwidth]{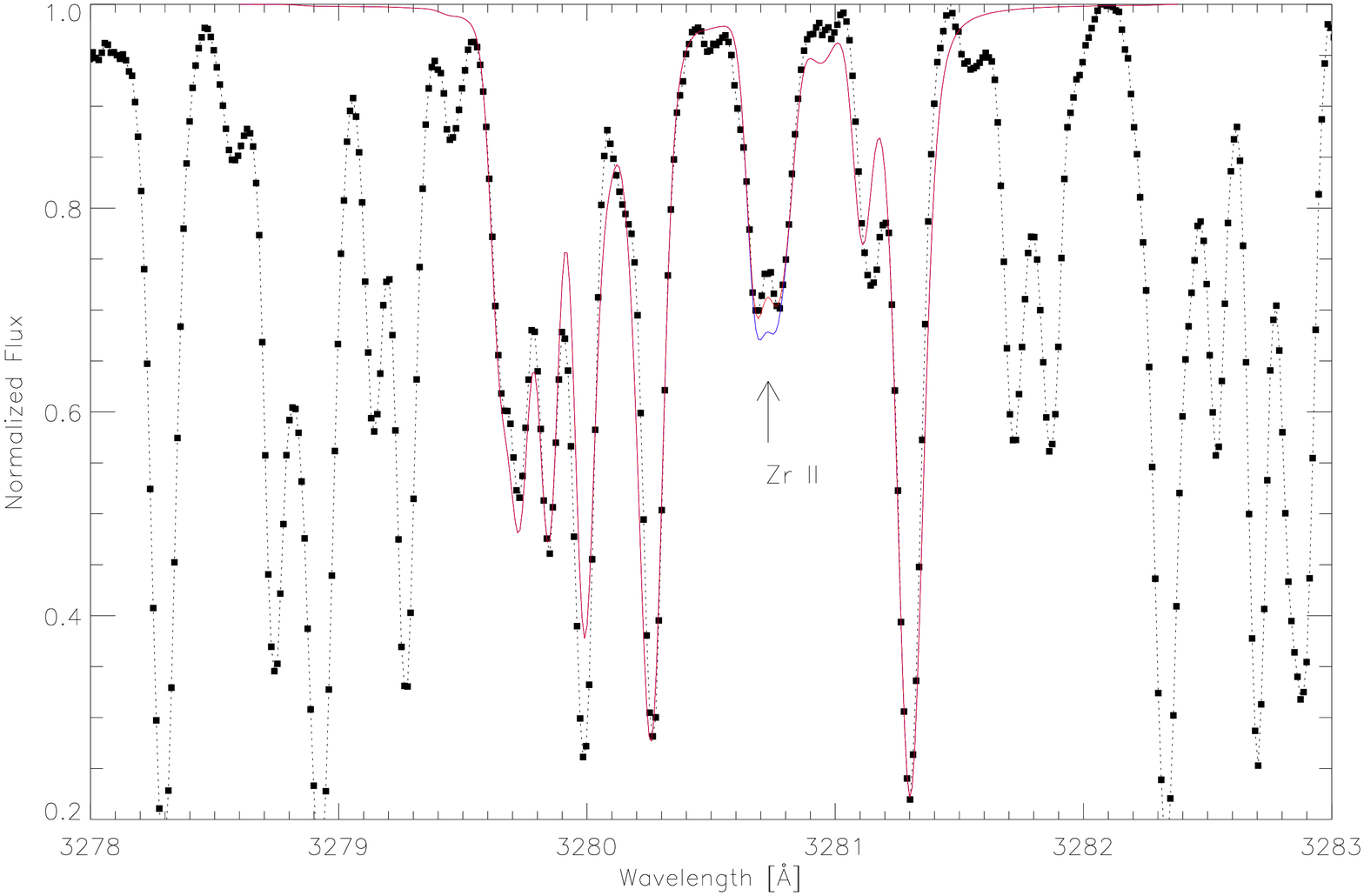}
    \vspace{-2mm}
    \caption{A five Angstrom part of the spectrum around the 3280\AA\, Ag line, which is blended with a Zr II and Fe I line.}
    \label{blends}    
    \end{center}
    \end{figure}

In order to derive the stellar abundances\footnote{For details on the sample and 1D LTE analysis see \pos{[8]}.} we need to fit line profiles to the spectral lines and determine the stellar parameters such as temperature, gravity and metallicity ([Fe/H]). As more remote, metal-poor stars are observed, the lines weaken or disappear from the stellar spectrum \pos{[5]}. The metallicity is therefore a direct measure for how many lines we expect to see. At the lowest metallicities we will only be able to detect and measure the strongest lines.
The metallicity is as all other '[A/B]' abundance notations an indicator of the number of absorbing atoms of species A with respect to B present in the outer part of the stellar atmosphere.
The metallicity, [Fe/H], of the star '*' with respect to the Sun '$\odot$' is defined as:
\begin{eqnarray}
[Fe/H] \equiv \log(N_{Fe}/N_{H})_* - \log(N_{Fe}/N_{H})_\odot
\end{eqnarray}
where each of the $\log(N)$ terms come from 
\begin{eqnarray}
\log W = \log(const) + \log(A) + \log(gf\lambda) - \theta \chi -\log(\kappa)
\end{eqnarray}
This shows how the abundance of element A, line oscillator strength ($\log gf$), temperature ($\theta$), excitation potential ($\chi$), and absorption coefficient ($\kappa$) are related to the equivalent width (W)\footnote{W corresponds to the integrated area of the spectral line.}. It is therefore extremely important that we know the atomic data, such as $\log gf$ and $\chi$, very accurately, both for the line but also for the surrounding lines. The heavy elements typically have their only lines in the near-UV part of the spectrum, where there are so many absorption lines, that these will blend into each other (see Fig. \ref{blends}). Therefore, we need to know exactly how much of the line profile belongs to the element of interest and to the blending lines. Hence, very accurate atomic information of blending lines is essential to derive their abundance as well, which is needed for an accurate spectral line and abundance analysis.

\section{The 'second' r-process}
\begin{figure}[!h]
    \begin{center}
    \includegraphics[width=0.38\textwidth]{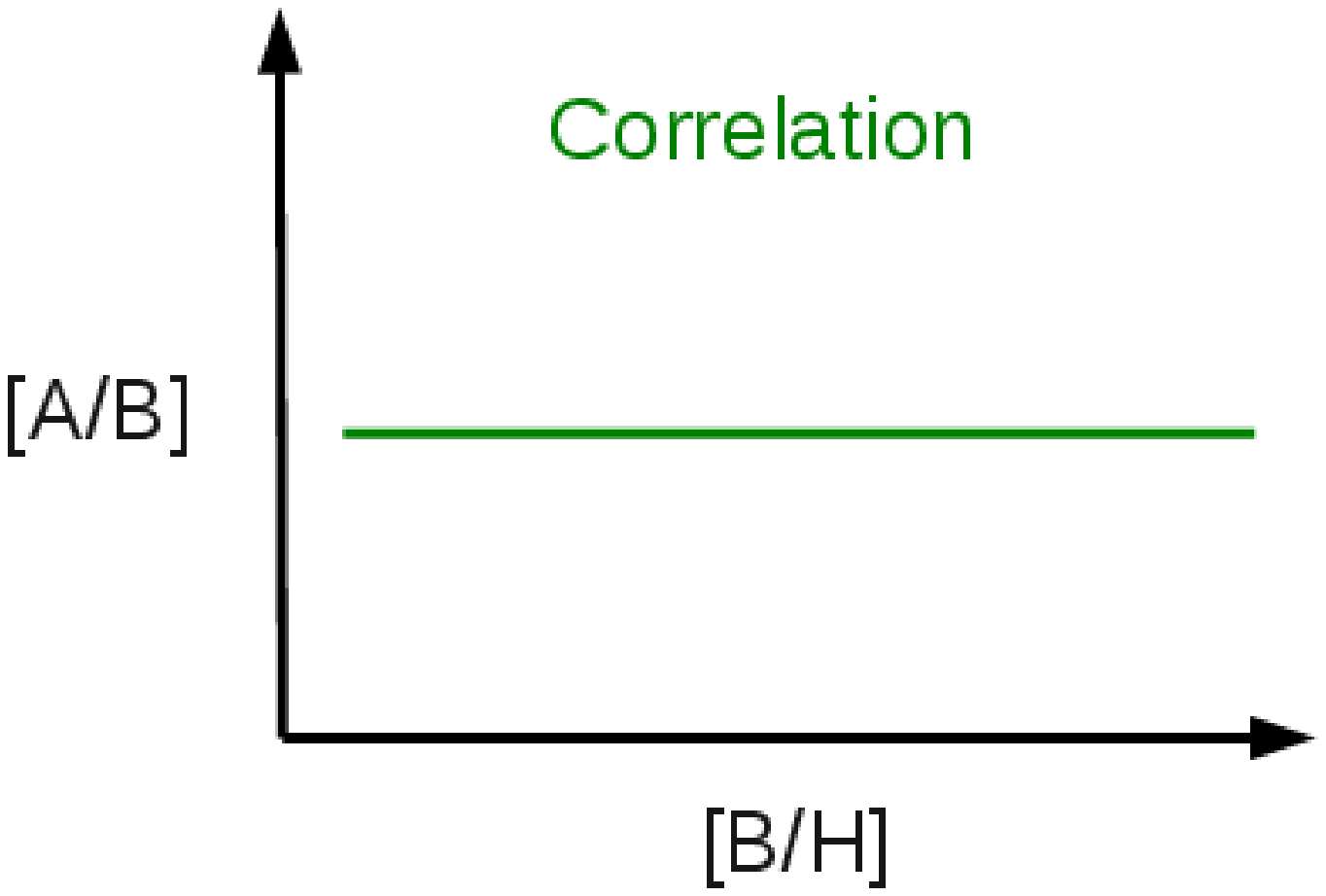} 
    \includegraphics[width=0.38\textwidth]{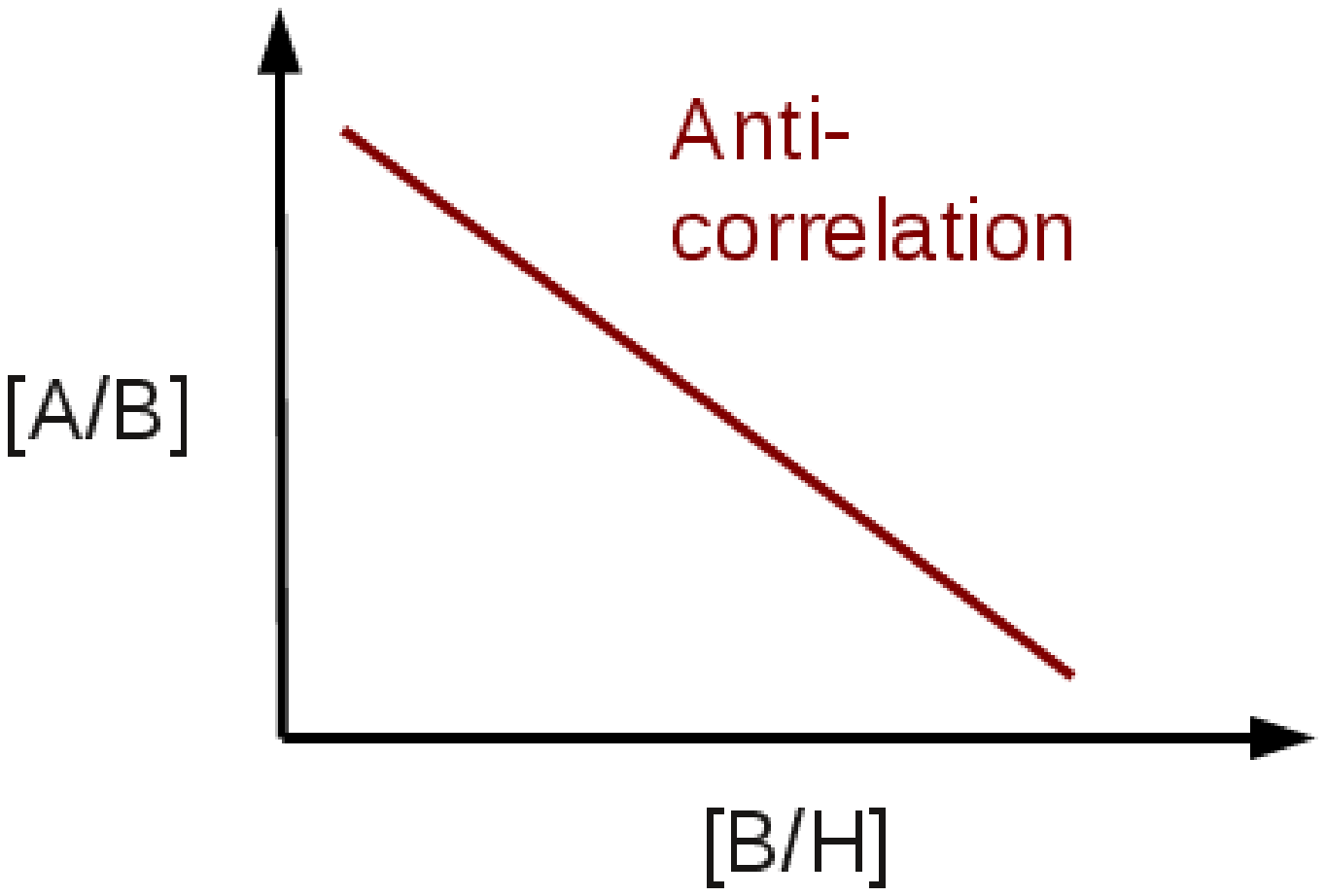}
    \vspace{-8mm}
    \caption{A correlation (left) and anti-correlation/'lack of correlation' (right) between element A and B.}
    \label{corr}
    \end{center}
    \end{figure}

To trace and understand the formation of Pd and Ag, five other heavy elements were studied: Sr, Y, Zr, Ba, and Eu. These are selected so that weak/main r-/s-processes are represented.
The weak s-process is traced by Sr and Y (and to some extend Zr \pos{[9]}). Zirconium, palladium, and silver are tracers of the weak r-process (see \pos{[8]}), while Ba is created by the main s-process, and Eu by the main r-process \pos{[2]}. Since the r-process is a primary process it can take place at the lowest metallicities independent of the presence of heavy seed nuclei, while the s-process, being a secondary process, needs such heavy seeds to build onto. Therefore, we will only see r-process elements at the very lowest metallicities, where heavy s-only elements were absent. Most of the above mentioned elements have a low fraction (10-20\%) that is created by the r-process\footnote{e.g. because their isotopes can be formed by both an r- and s-process, and the r-process will be the dominant at low [Fe/H] i.e. in the early Galaxy.}. 
Strontium is created by the weak s-process (85\% \pos{[2]}) or the charged particle process at low [Fe/H]. This element shows, as most other heavy elements, a very large star-to-star scatter ($>2$dex), which is by far in excess of any observational uncertainty or model assumption \pos{[1,7]}.
This scatter indicates that tracing the formation process is not straightforward. There might be several (different) formation processes, that contribute to the abundances that we derive, and the efficiency of each process might be far from uniform.
\begin{figure}[!h]
    \begin{center}
    \includegraphics[width=0.65\textwidth]{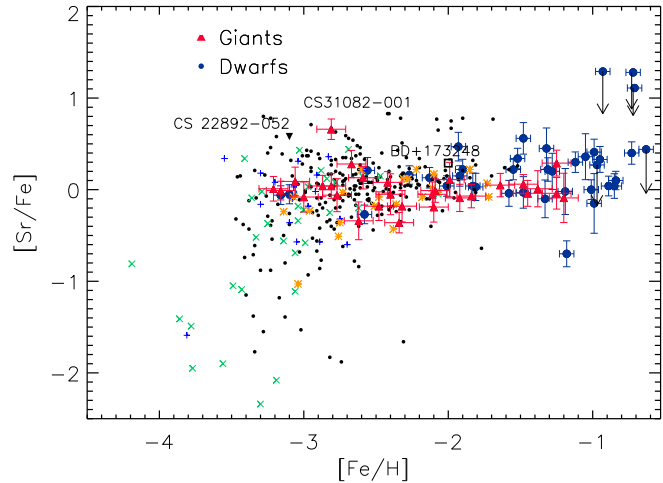}
   \vspace{-2mm}
    \caption{[Sr/Fe] as a function of [Fe/H] for dwarfs (blue circles) and giants (red triangles). Details on the comparison samples are given in [8] and references therein.}    
    \end{center}
    \end{figure}
By comparing elements for which we know the dominant production processes, we can investigate the elements' origin and evolution, to see how these agree or differ. This will be expressed through abundance correlations and anti-correlations. Take element A (weak s-process created) and compare it to another weak s-process element (B). They will show a correlation, because they are created by the same process, that equally efficiently might create both elements (see Fig. \ref{corr}, left figure). However, if B is instead created by a main r-process, these two elements will most likely grow independently of each other, and at different rate, creating an anti-correlation (Fig. \ref{corr}, right hand plot).
\begin{figure}[!h]
    \begin{center}
     \includegraphics[width=0.5\textwidth]{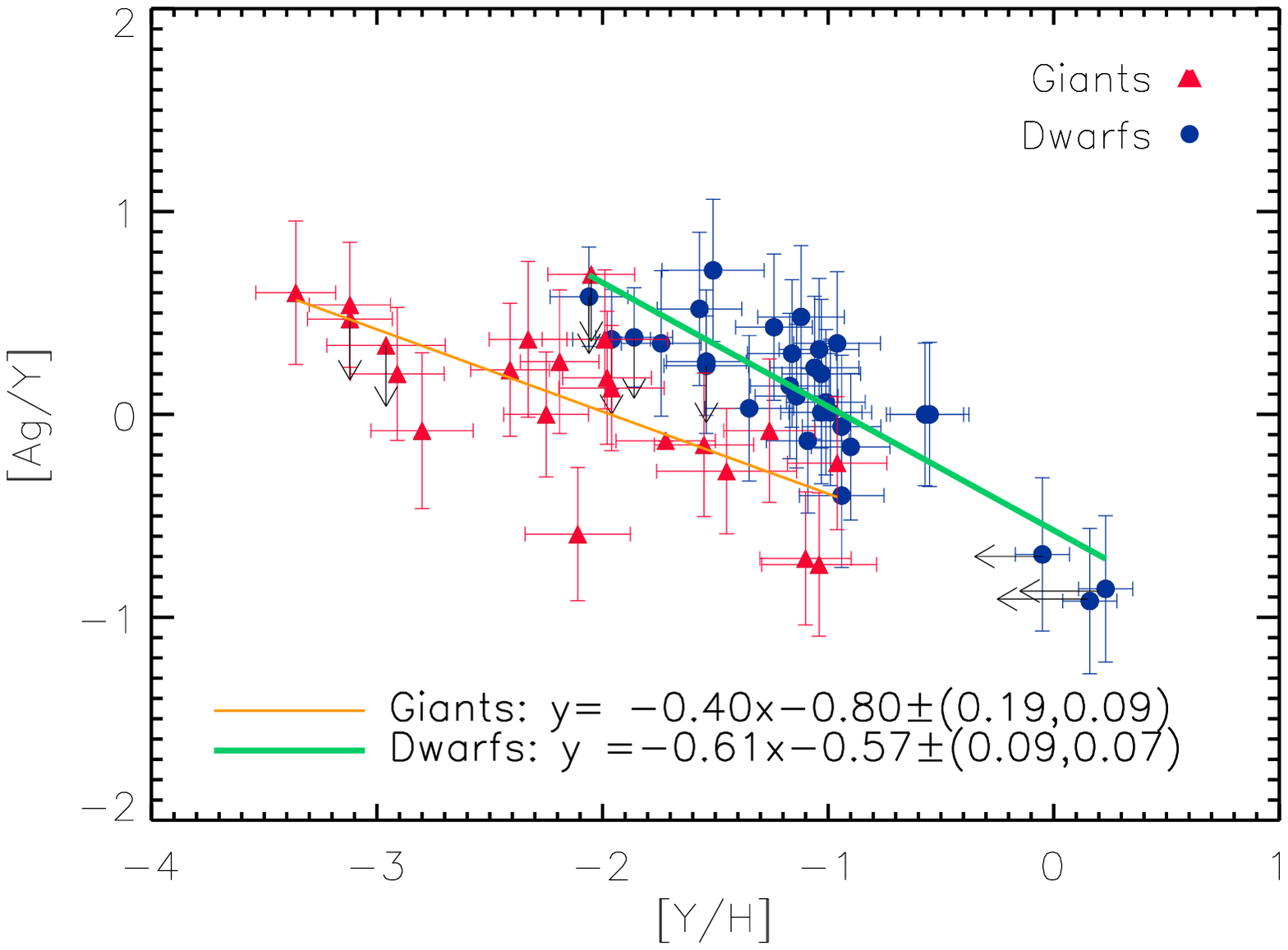}
    \vspace{-5mm}
     \includegraphics[width=0.49\textwidth]{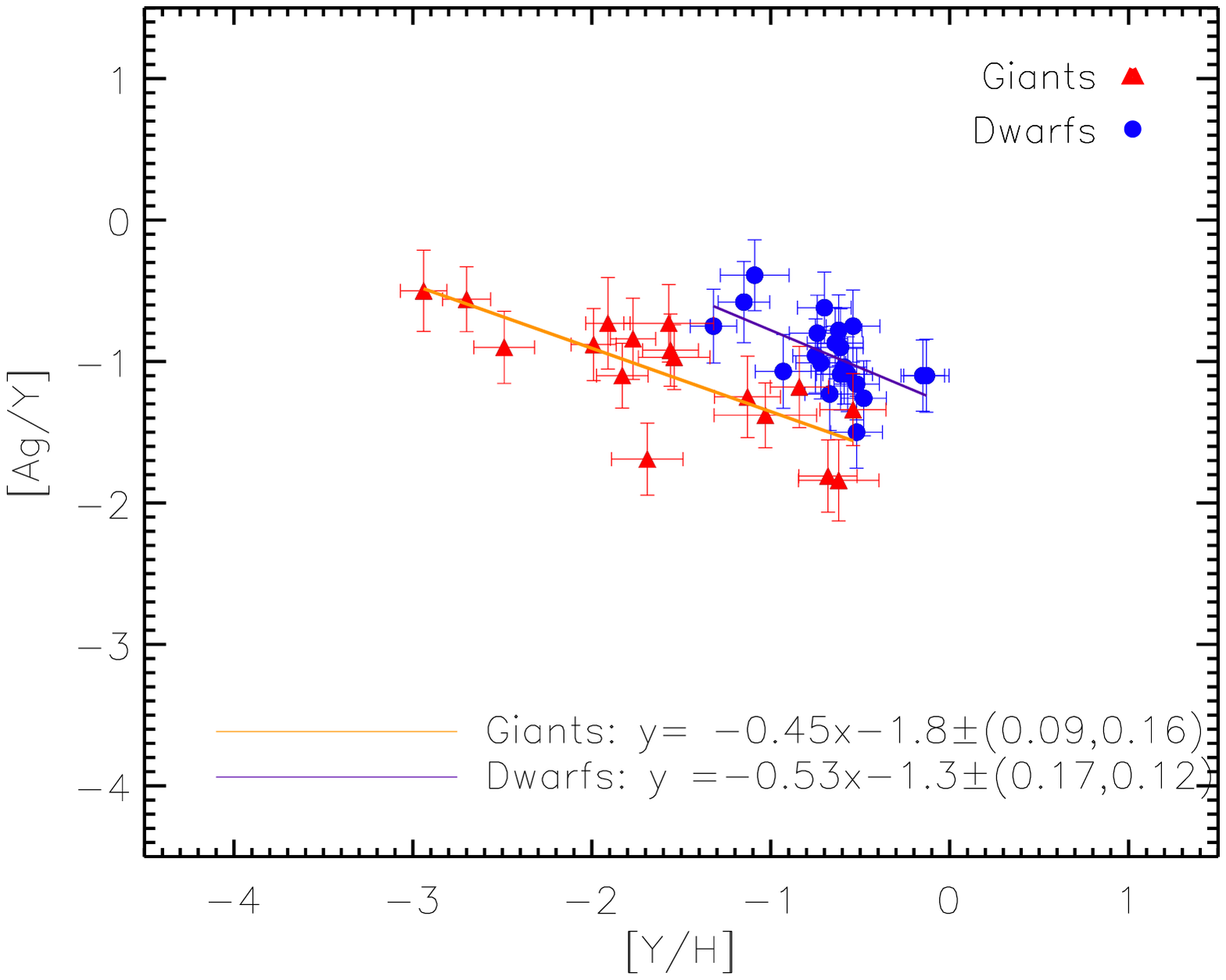}
     \includegraphics[width=0.49\textwidth]{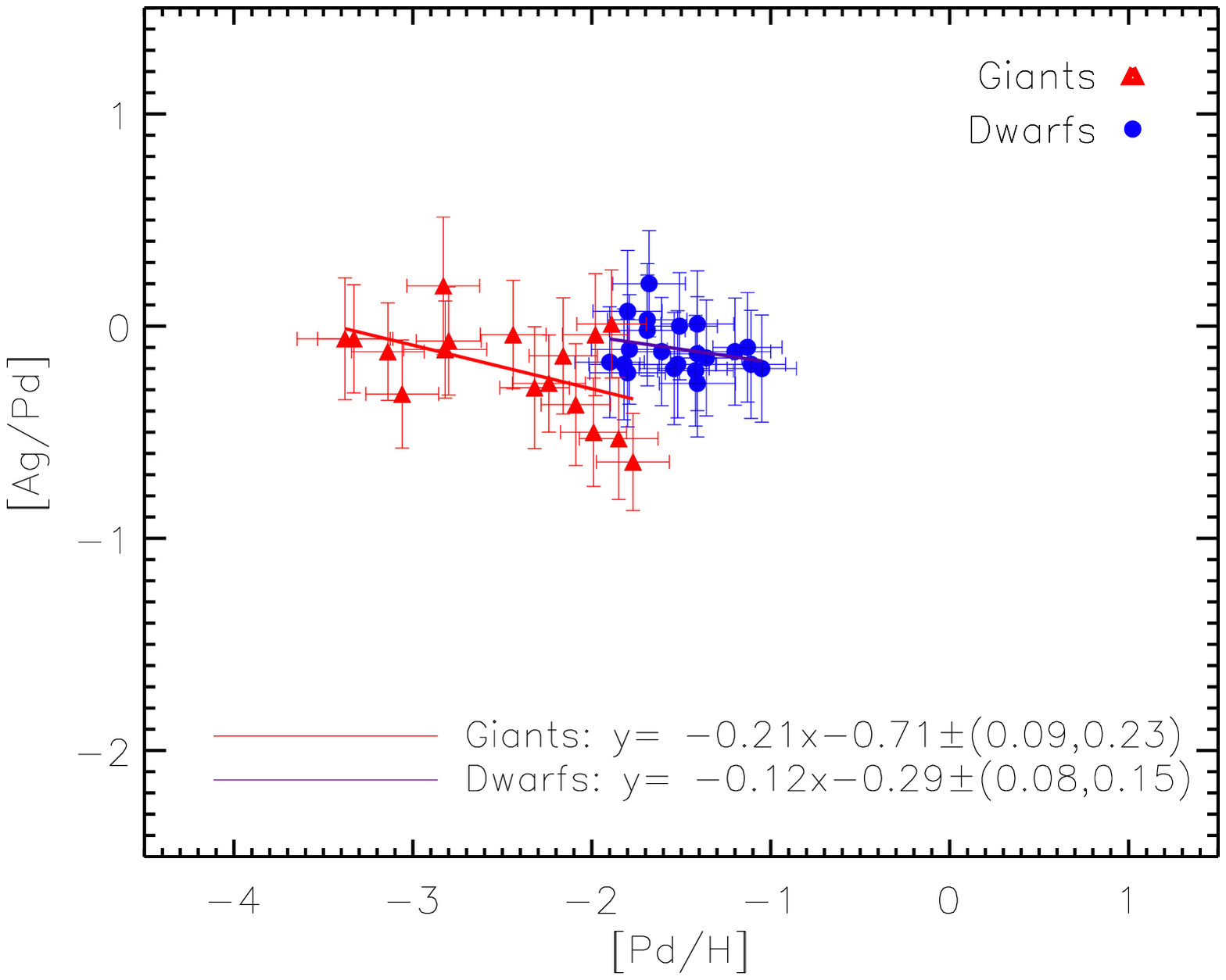}
    \includegraphics[width=0.49\textwidth]{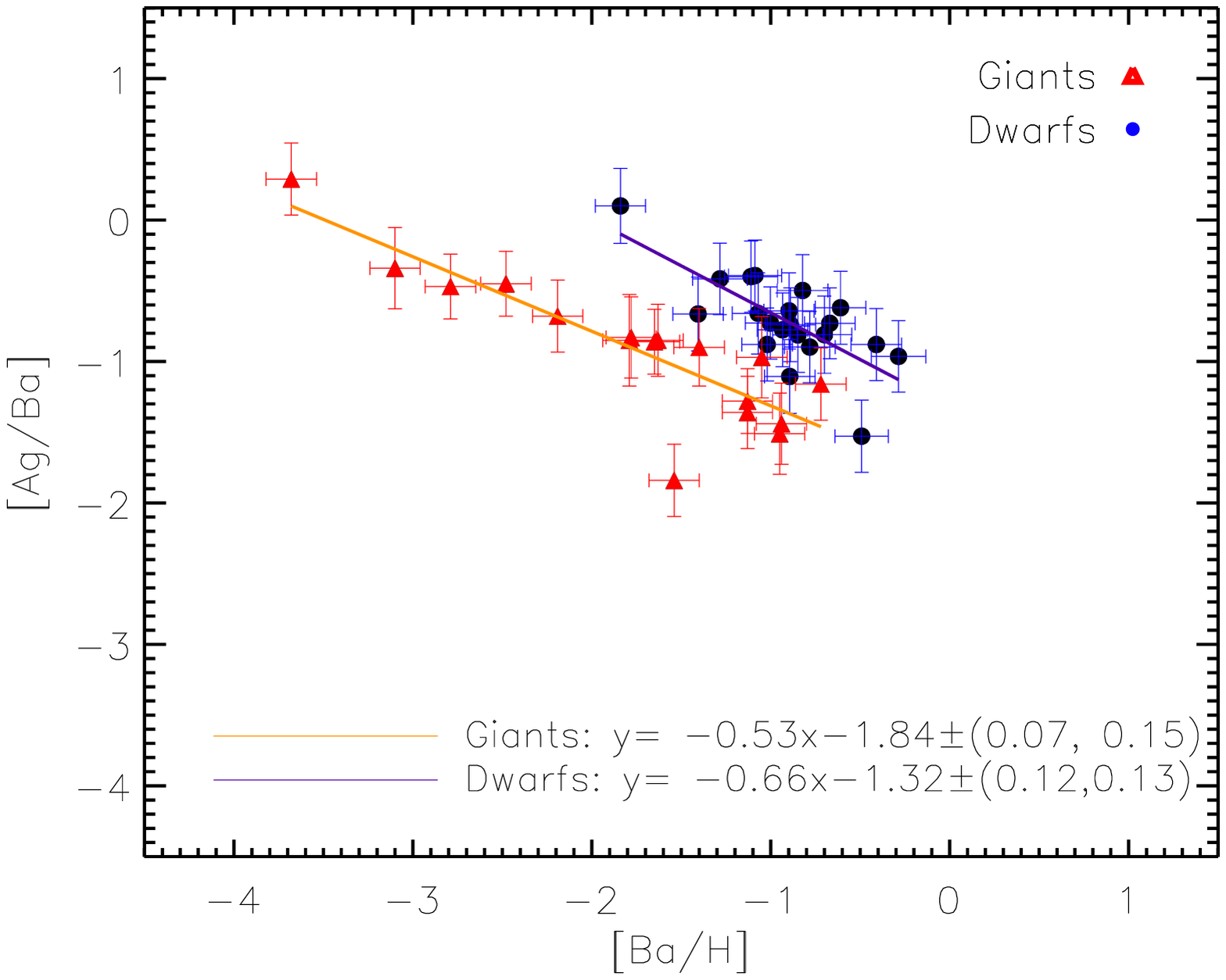}
\vspace{-2mm}
    \caption{Correlations and anti-correlations of Ag with Y, Pd and Ba. The upper left figure includes upper limits while the other figures do not.}
    \label{YPdBa}
    \end{center}
    \end{figure}

Figure \ref{YPdBa} shows a weak s-process element, Y (92\% - \pos{[2,9]}), a weak r-process element, Pd (54\%), and a main s-process element, Ba (81\%), which may be created by the main r-process at low metallicity.
The two upper plots in Fig. \ref{YPdBa} both show [Ag/Y] vs [Y/H], where the figure to the left includes the upper limits, while the one to the right does not. The slope of the line fitted to the dwarfs decreases and becomes almost parallel with the line fitted to the giants\footnote{The sample is slightly biased towards extremely metal-poor giants, i.e no giant has [Fe/H] $>-1$} when disregarding upper limits. The overall trend and anti-correlation remains the same with or without upper limits. However, the uncertainty related to the fit grows with decreasing number of measurements. Several different fitting techniques were applied, but the slopes are robust and change on average by $\pm0.05$. The errors on derived abundances are taken into account in the fit.
Furthermore, the dwarfs and giants are fitted separately, so that possible systematics can be traced. A possible explanation for the shift of the line fitted to the dwarfs and giants, respectively, could be due to NLTE or 3D effects, which to date have not been determined for Ag or Pd. 

Silver is clearly seen to anti-correlate with both Y and Ba at all metallicities (Fig. \ref{YPdBa}). Hence, Ag is not created by a weak or a main s-process, and also not by a main r-process (since Ag and Ba also anti-correlate at low metallicity). This is confirmed by anti-correlation of Ag with Sr and Eu (see \pos{[8]}). Silver and palladium correlate indicating that they share the same formation process. 
Therefore, Ag and Pd must be created by a weak/second r-process, a process that is clearly different from the main r-/s-process as well as the weak s-process. 
Comparing the abundances (Sr-Eu) from each star to yield predictions we can loosely estimate the characteristics of the weak r-process. A lower neutron density and entropy are needed for this weak r-process compared to the main r-process \pos{[6]}, and an electron fraction of $0.2-0.3$ seems necessary in this scenario. This interval is estimated from a comparison of abundances to the yields from \pos{[11]}, and it is higher than what is needed for a main r-process, but lower than what an s-process requires.

\section{Neutron-capture elements in Surveys}
\begin{figure}[!h]
    \begin{center}
    \includegraphics[width=0.56\textwidth]{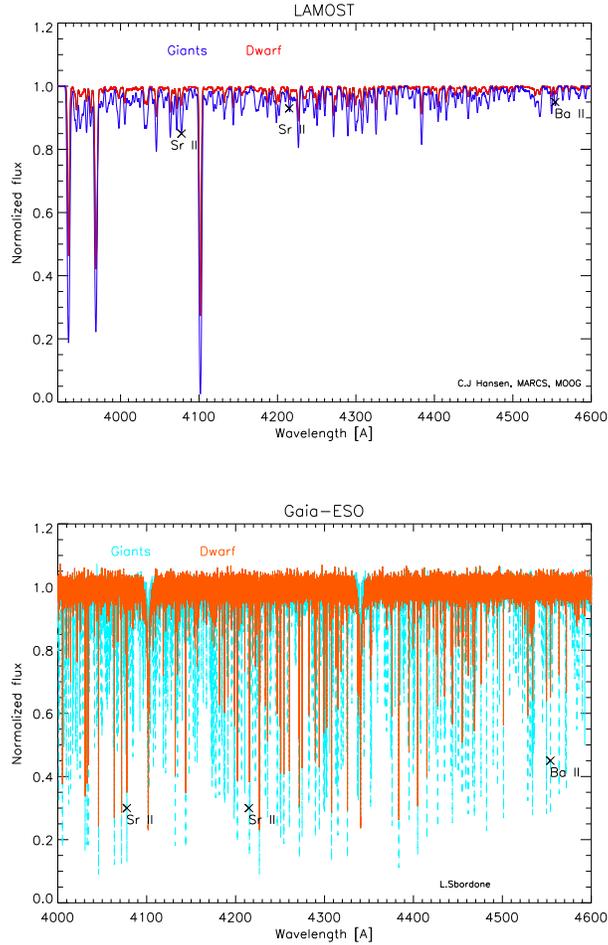}
   \vspace{-2mm}
    \caption{Normalized flux versus wavelength for a synthetic UVES spectrum with noise injected to simulate S/N=50. The giants are shown in blue colors and the dwarfs in red.} 
    \label{survey}
    \end{center}
    \end{figure}

Only Sr (and Ba) might be detectable in both low resolution metal-poor stars observed with
LAMOST (low resolution, $R \sim 1800$) and also with UVES/VLT (high resolution, $R \sim 40000$). With UVES many heavy elements can be measured, so the resolution and wavelength range are limiting factors.
Despite the fact that most surveys focus on more red regions (such as 480-680nm from Gaia-ESO using UVES), we can use the information from these spectra to select follow-up candidates to re-observe in the near-UV. The quality of spectra that we expect to gain from LAMOST and UVES is shown in Fig. \ref{survey}. 
\begin{figure}[!h]
    \begin{center}
    \includegraphics[width=0.56\textwidth]{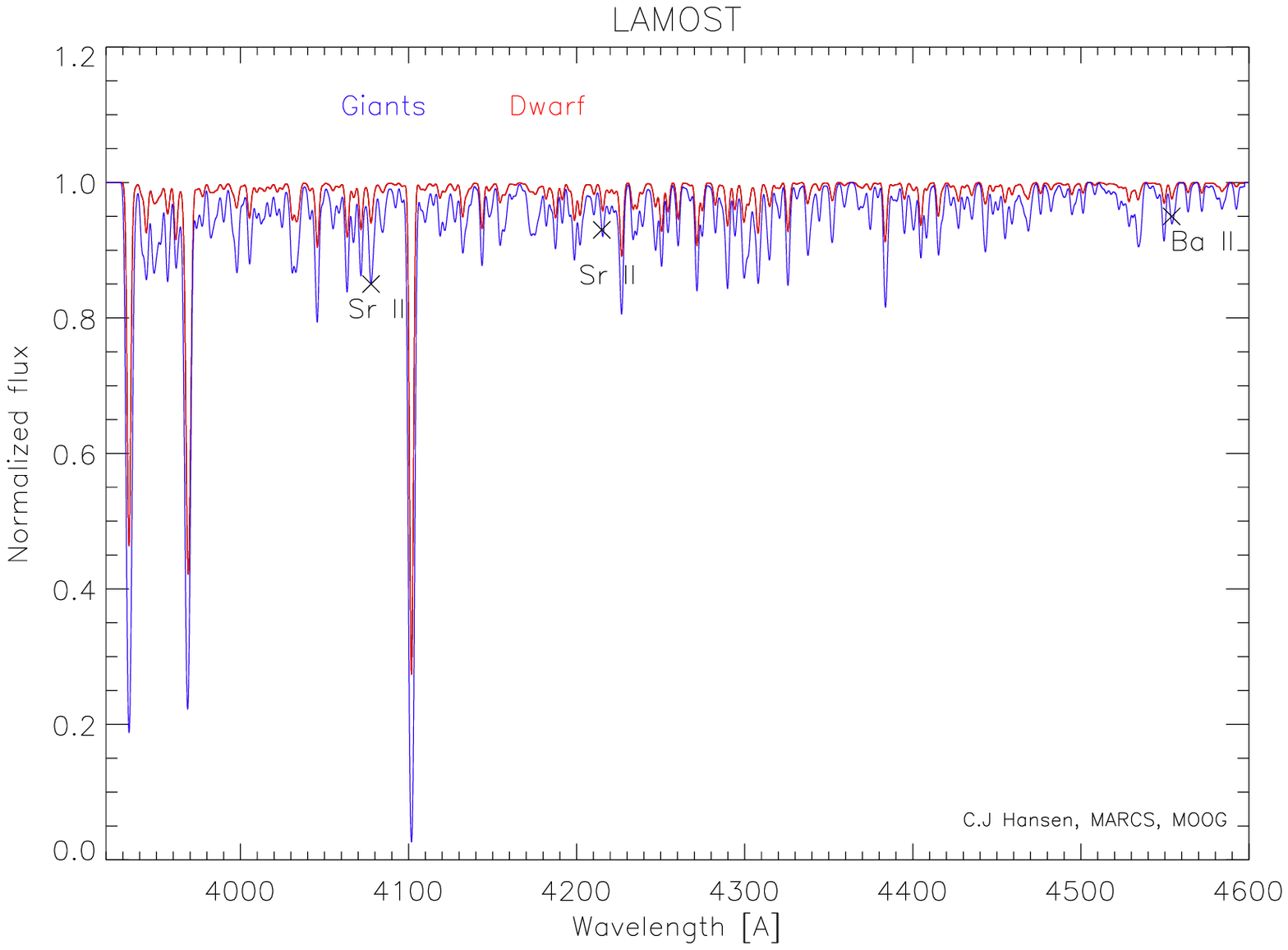}
   \end{center}
    \end{figure}
\vspace{-10mm}
Figure \ref{survey} shows part of the blue region, where the majority of the neutron-capture elements have their lines. The strongest Sr and Ba lines have been marked with crosses. Only in high signal to noise (S/N) LAMOST spectra of neutron-capture enriched stars we might be able to detect Sr in metal-poor giants. Furthermore, this is only possible if the S/N$>50$\footnote{S/N estimated by comparing the noise band from the UVES spectrum to the depth of the 4077\AA\, Sr line in the LAMOST spectrum.}.

The large flow of data from current surveys makes it increasingly important to derive the stellar abundances as accurately as possible, which is why we need to know the impact of e.g. non local thermodynamic equilibrium (NLTE) effects on the Sr abundances. This is necessary to interpret the chemical evolution of these heavy elements correctly.

\subsection{The importance of NLTE for strontium}
\begin{figure}[!h]
    \begin{center}
    \includegraphics[width=0.45\textwidth]{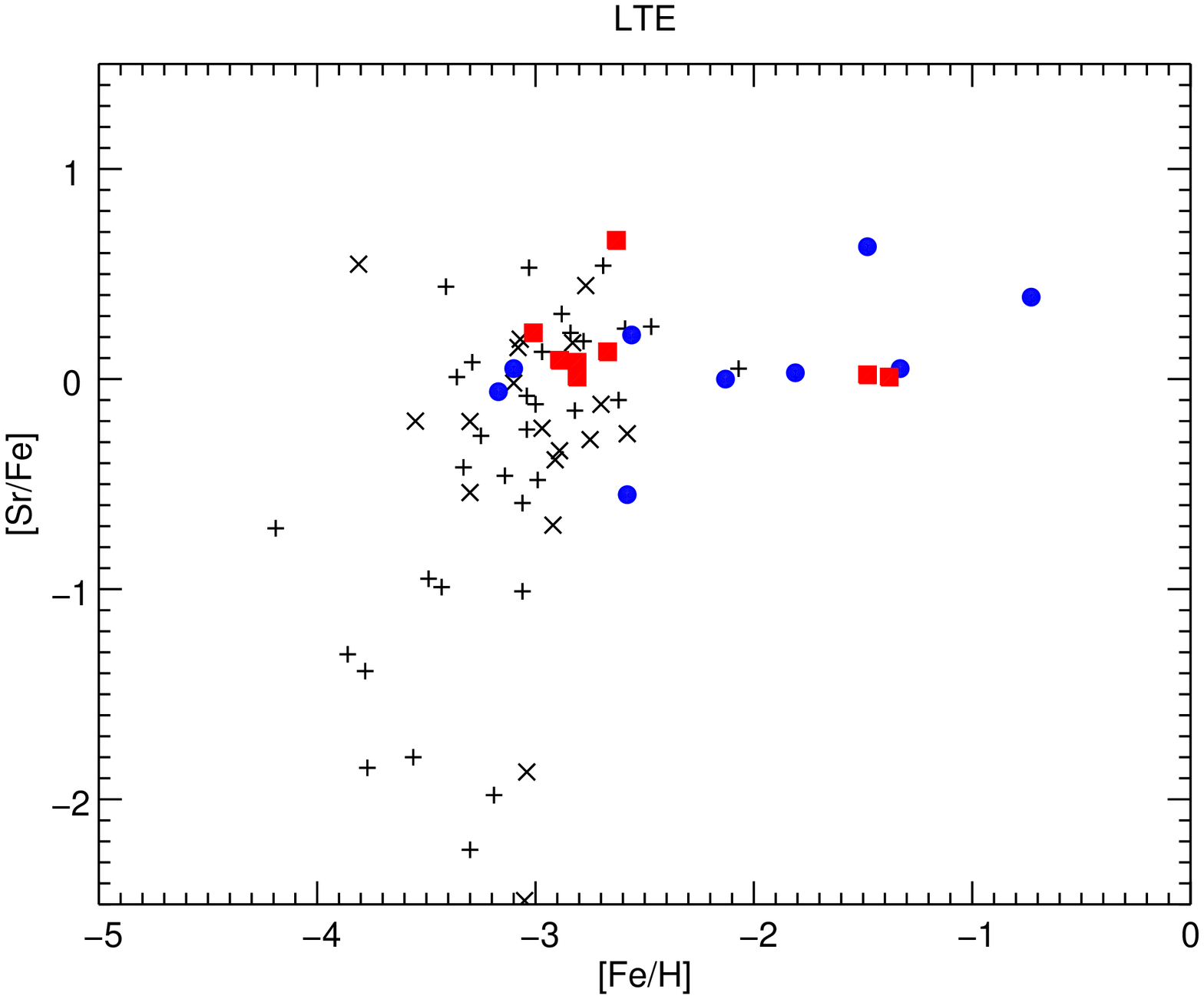}
    \includegraphics[width=0.45\textwidth]{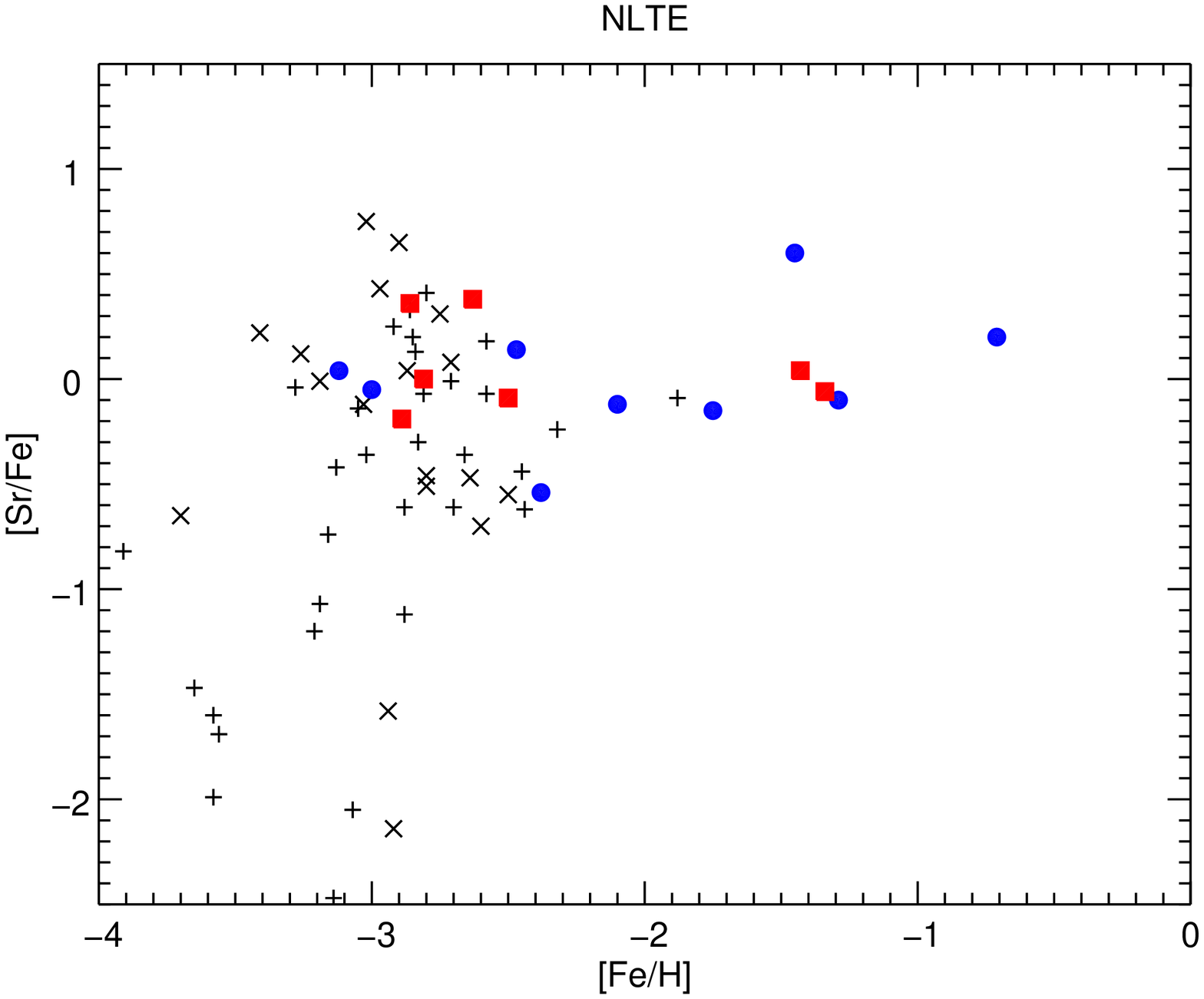}
   \vspace{-2mm}   
    \caption{[Sr/Fe] vs [Fe/H] derived under LTE (left) and NLTE (right). The red squares/blue circles represent our giants/dwarfs, while pluses/crosses represent the First Stars giants/dwarfs, respectively.}
   \label{ltenlte}
    \end{center}
    \end{figure}

As outlined above, Sr will be a key neutron-capture elements, if not the only one, we can measure in low resolution spectra. Figure \ref{ltenlte} shows a sample from Hansen et al, 2012b (submitted) of dwarfs and giants compared to the First Stars sample of dwarfs from \pos{[4]} and the sample of giants from \pos{[7]}. 
Stellar parameters as well as abundances have been corrected for NLTE effects \pos{[1,3]}. The stellar parameters were corrected by applying NLTE corrections to the Fe abundances, and the gravities were adjusted by using the approximate [Fe/H] --- log g scaling from \pos{[10]}. The Sr NLTE corrections computed with the model atom described in \pos{[3]}. Strontium shows a large star-to-star scatter, which is far in excess of any observational uncertainty. This scatter remains under NLTE. {\it The chemical evolution of Sr (down to [Fe/H]=$-3$) is the same under both LTE and NLTE, and it is safe not to apply NLTE corrections (if these are not available). However, below this metallicity NLTE corrections ought not to be neglected} (Hansen et al, 2012 submitted).
\begin{figure}[!h]
    \begin{center}
    \includegraphics[width=0.25\textwidth, angle=-90]{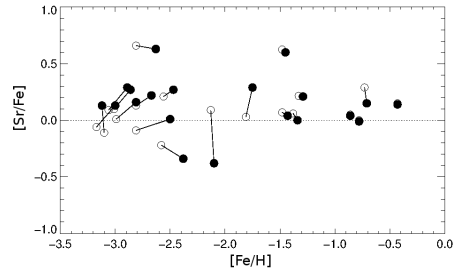}
    \caption{LTE Sr abundances (open circles) and the NLTE corrected Sr, derived with NLTE corrected stellar parameters (filled circles) for our sample.}
    \label{diffnlte}
    \end{center}
    \end{figure}
Figure \ref{diffnlte} shows the difference of LTE abundances based on LTE derived stellar parameters versus the NLTE derived Sr abundances based on NLTE corrected stellar parameters. The main change in Sr$_{LTE}$ vs Sr$_{NLTE}$ abundances stems from the NLTE corrected stellar parameters, since the Sr II NLTE corrections are on average small ($\sim 0.05$dex).

\section{Summary}
When using abundances to draw conclusions on yields and Galactic chemical evolution, it is important that we know them as precisely as possible. Therefore, 3D and NLTE effects should be considered when they are available. This is unfortunately not always possible either due to lacking atomic data or model atoms. The impact of NLTE corrected Sr abundances is minor in scheme of GCE modeling. This is mainly due to the fact that these models are calculated with very uncertain yield predictions, and because Sr shows such a large star-to-star scatter. However, the LTE Sr II abundances for [Fe/H]$>-3$ are reliable.
This gives strength to LTE abundances derived from ionized majority species. The correlation between Ag and Pd seems robust, as do their anti-correlations with Sr, Y, Zr, Ba, and Eu. These correlations and anti-correlations strongly indicate the need for a second/weak r-process, which would create neutron-capture elements such as Pd and Ag, but not the five other elements studied. The offset between the lines fitted to dwarfs and giants, respectively, might be explained by NLTE and 3D effects, which we need to know for all seven elements before we can NLTE correct the abundances in the [A/B]-ratios.

\end{document}